\documentclass{article}
\usepackage{graphics}
\begin{document}

\begin{center}
{\large\bf Numerical studies of Anderson transition}

{P.\ Marko\v s}\\
 
  {Institute of Physics, Slovak Academy of Sciences,
  D\'ubravsk\'a  cesta 9, 842 28 Bratislava, Slovakia}
\end{center}
\begin{abstract}%
We present  numerical  results for the
statistics of $z$'s 
($z$'s are defined as logarithm of  eigenvalues of the transfermatrix $T^\dag T$)
at the  critical points of Anderson transition   in 3D and 4D.
The  change of the density of $z$ due to the crossover from the metallic
to  the localized  regime is described. 
Linear behavior  
$\rho(z)= z$ at the critical point in 3D is proven and discussed. In the insulating regime, 
the universal form of $\rho$ has been found.
\end{abstract}

~~~~~~\\
While the finite-size scaling analysis of the disorder-induced
metal-insulator transition enables us to find critical disorder $W_c$
and critical exponent $\nu$ 
from the knowledge of the smallest positive Lyapunov exponent, $z_1$
in the quasi-one dimensional limit
\cite{MacK},
the knowledge of the 
higher $z$'s are necessary to understand the  statistical properties 
of transport in cubic samples.
Here, $z_i$ is defined as logarithm of the $i$-th eigenvalue of the matrix
$T^\dag T$ ($T$ is the transfer matrix). 

Statistics of $z$'s is well known understood in the 
limit of the  small disorder. Its  properties, derived in the analogy with
the random matrix theory,
\cite{pichardnato}, 
enables us to explain universal features of the transport
in weakly disordered mesoscopic systems
\cite{beenak}.
The key role in this explanation plays the density $\rho(z)$, defined as
\begin{equation}\label{rho}
\rho(z)=\langle\sum_i\delta(z-z_i)\rangle,
\end{equation}
where the  summation covers all channels and brackets mean averaging over
statistical ensemble. In the weak disorder limit, $\rho(z)$ = const.
Another typical characteristics of the distribution $(\delta)$ of (normalized)
differences between $z$'s. In the weak disorder limit, $P(\delta)$
equals to the Wigner surmise $P(s)=\frac{\pi}{2}s\exp-\frac{\pi}{4}s^2$
\cite{pichardnato}.

Our aim in this paper is to shown how the statistics of $z$'s changes
when system undergoes the metal-insulator transition. We believe that the
understanding
of the statistical properties of $z$'s at the critical region would
provide us with serious basis for the more general understanding  of the Anderson
transition, including  the description of the system size and disorder
dependence of conductance and its statistical moments in the critical region.

We consider Anderson model: 
\begin{equation}\label{ham}
{\cal H}=W\sum_n\varepsilon_n|n\rangle\langle n|+\sum_{\rm [n.n.]}|n\rangle\langle n'|.
\end{equation}
In (\ref{ham}), $n$ numerates sites on the
$d$-dimensional cubic lattice $L^d$, 
and $W$ measures the disorder. 
For the box distributed random energies $\varepsilon$,
$|\varepsilon|<1/2$, model exhibits the metal-insulator transition
at the critical point $W_c\approx 16.5$ (34.5) for $d=3$ (4), respectively.

Recently, 
\cite{muttalib,jpcm} some speculative models has been proposed, in which
weak disorder  statistics of $z$'s has been generalized to models which  reflect
some special features of the metal-insulator transition.  Unfortunately, no one 
of them 
succeeded to describe the metal-insulator transition completely.


The analysis of the statistics of $z$'s has its counterpart in the level
statistics
\cite{shkl}. Here, the most important quantity is  the distribution $P(s)$
of the (normalized) differences between energy levels. It has been shown,
that there are three typical form of $P(s)$: Wigner surmises (WS) for the metallic, Poison for the insulating and the third, universal distribution at the critical point. 
The same scenario has been found for the statistics $P(\delta)$ of the
(normalized) differences between $z$'s
\cite{pm}.


We present in Fig. 1
$P(\delta)$ in the critical point. 
Data  show that (i) $P(\delta)$ is system size invariant,
(ii) it depends on the dimension,  and that 
(iii) it 
follows neither the Wigner surmises
nor the Semipoison distribution $P(s)=4s\exp(-2s)$.

\begin{figure}[t]
\centerline{\resizebox{7.1cm}{7.1cm}{\includegraphics{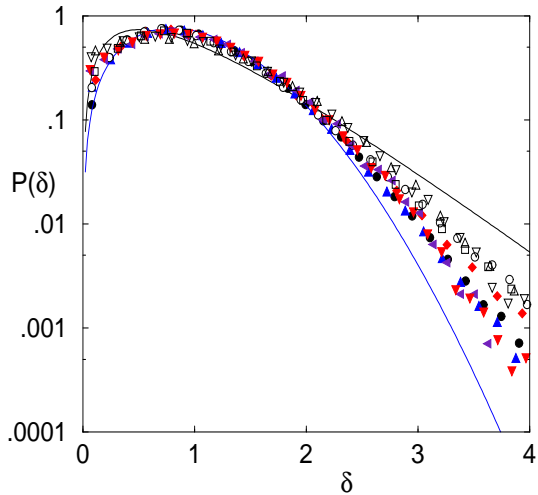}}}
\vspace*{-4mm}
 \caption{$P(\delta)$ for 3D (full symbols) and 4D (open symbols)
at the critical point. Solid lines are Wigner surmise and Semipoisson
distribution.}
 \label{delta}
\end{figure}

The first too  results correspond to that of the level statistics; the second one
confirms that the statistics of $z$'s is more similar  to the localized one in 
higher dimension
\cite{zk}. 
However, the third one is in the contradiction with   numerical 
observation  reported in Ref.
\cite{mont}, where
the Semipoisson distribution of the level statistics has been found as the result
of the sensitivity of the level statistics to the boundary conditions at the critical
point.  Although we consider only periodic boundary condition in this work,
we do not believe that the use of different boundaries will 
remove this disagreement. 
The reason is that the typical values of $z$'s  are rather large
at the critical point:  The    distribution of 
higher $z$'s is Gaussian  with mean value $\ge 5$ and variance $\sim 1$
\cite{pm}.
It is therefore highly unprobale that the change of boundaries could influence the statistics of higher $z$. In other words, higher channels are strongly localized (although 
mutually correlated) and so they do not feel the form of the boundaries. 


\begin{figure}[t]
\centerline{\resizebox{7.1cm}{7.1cm}{\includegraphics{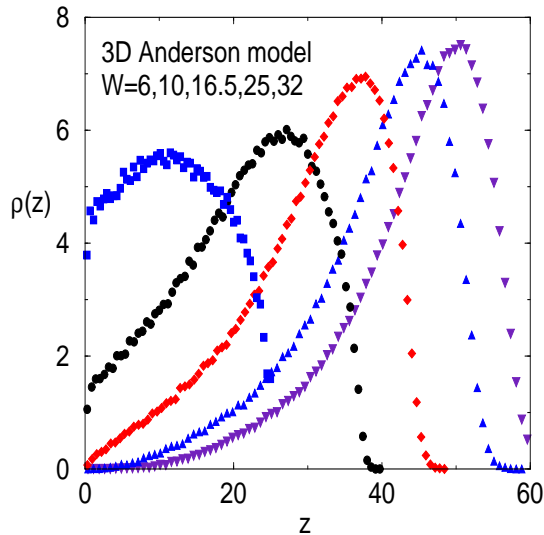}}}
 \caption{The density $\rho(z)$ for cubic system with  $L=12$.}
\end{figure}

Fig. 2.  shows how  the density  $\rho(z)$
changes due to the growth of disorder. 
The typical semicircle form of the $\rho(z)$ could be found only for
extremely small disorder ($W=6$) (we remove the contribution of closed channels
in this case).
The growth of the disorder deformates $\rho(z)$:  
the last decreases for small  $z$ in favor of the maximum, which moves 
towards the higher values of $z$. At the critical point, $\rho(z)$
becomes linear (quadratic) in $z$ in 3D (4D), respectively. 
This agrees with our previous result,
obtained for the quasi-one dimensional samples
\cite{jpcm}. 

\begin{figure}[t]
\centerline{\resizebox{13cm}{4.25cm}{\includegraphics{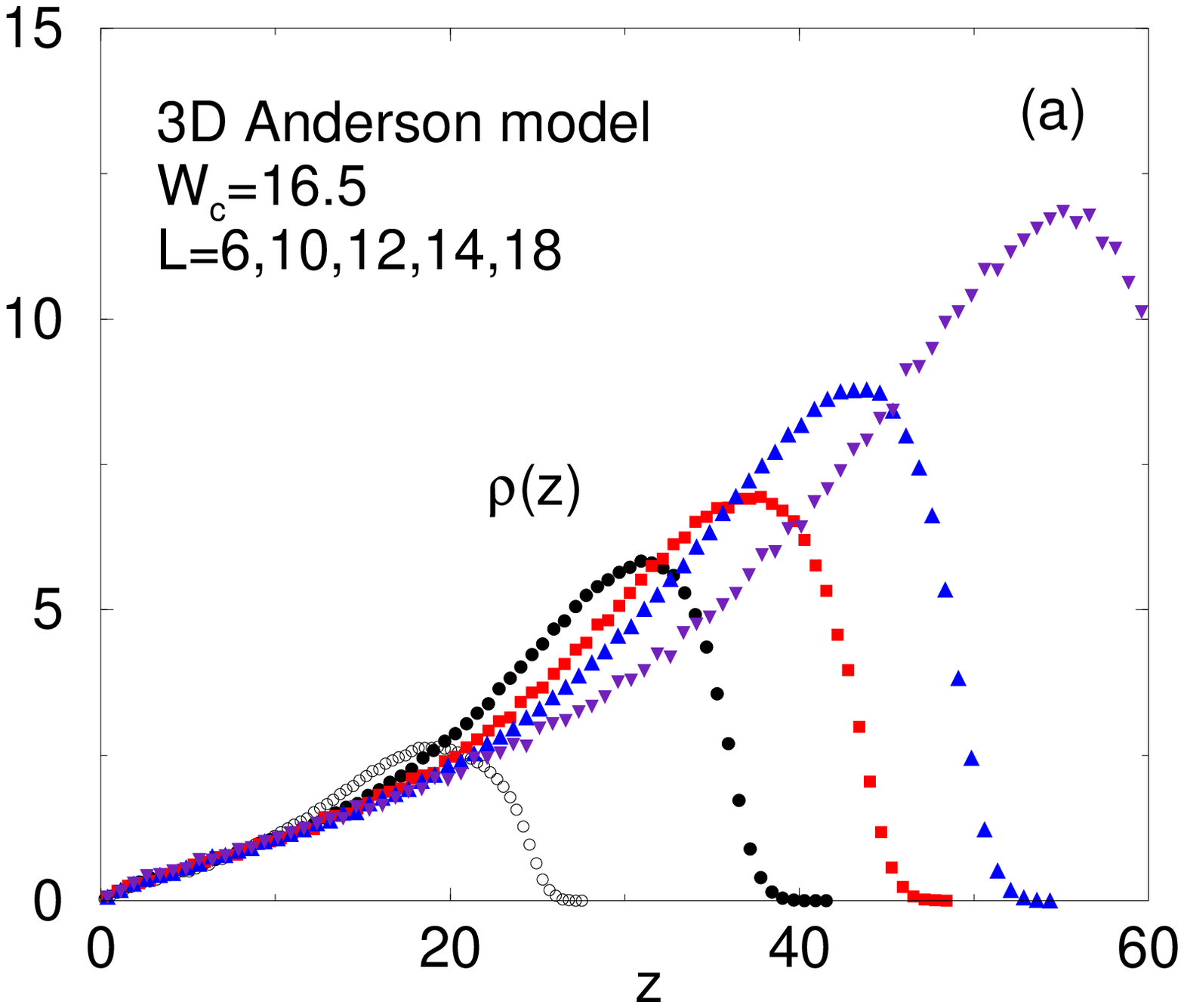}~~~~
\includegraphics{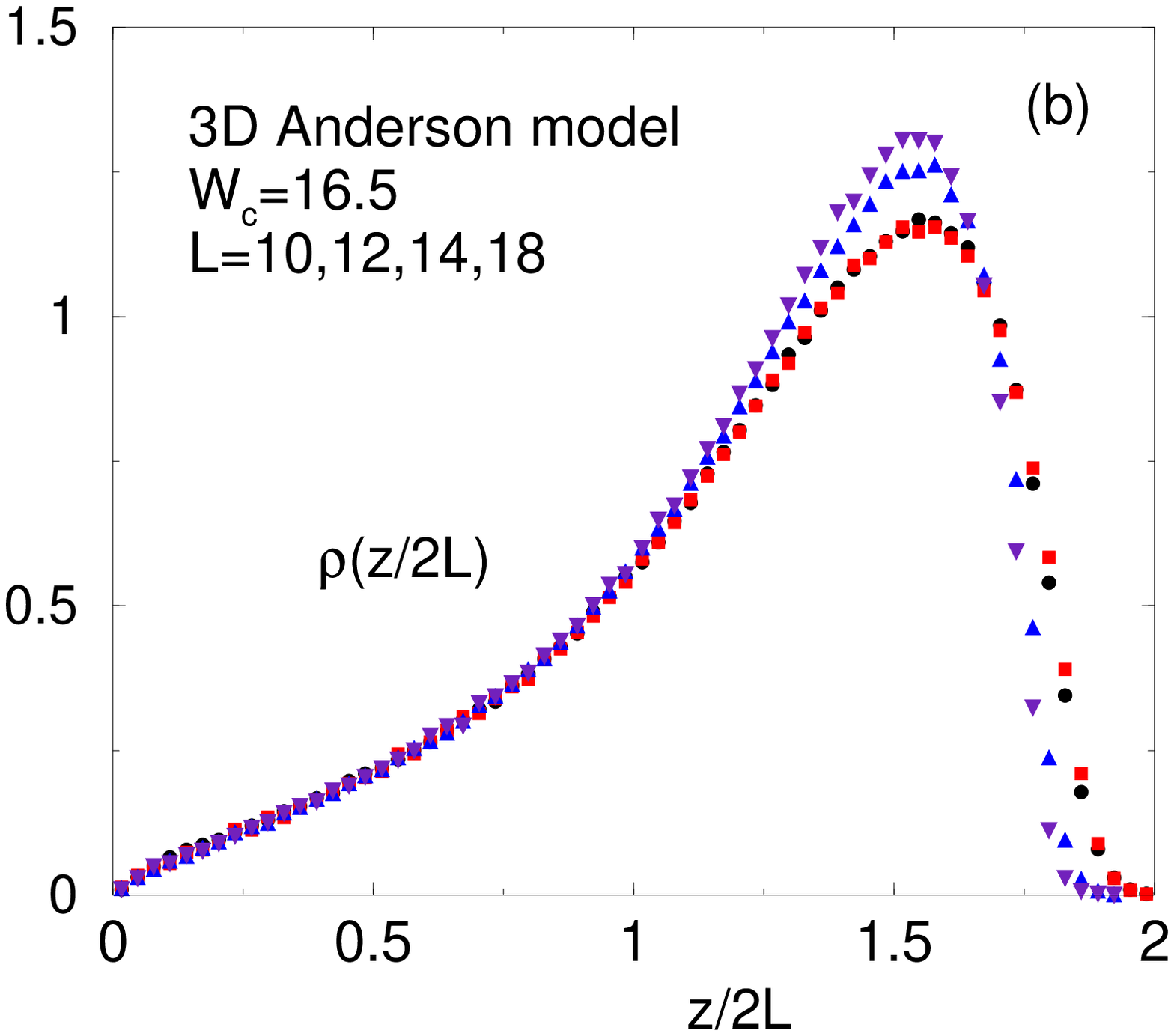}~~~
\includegraphics{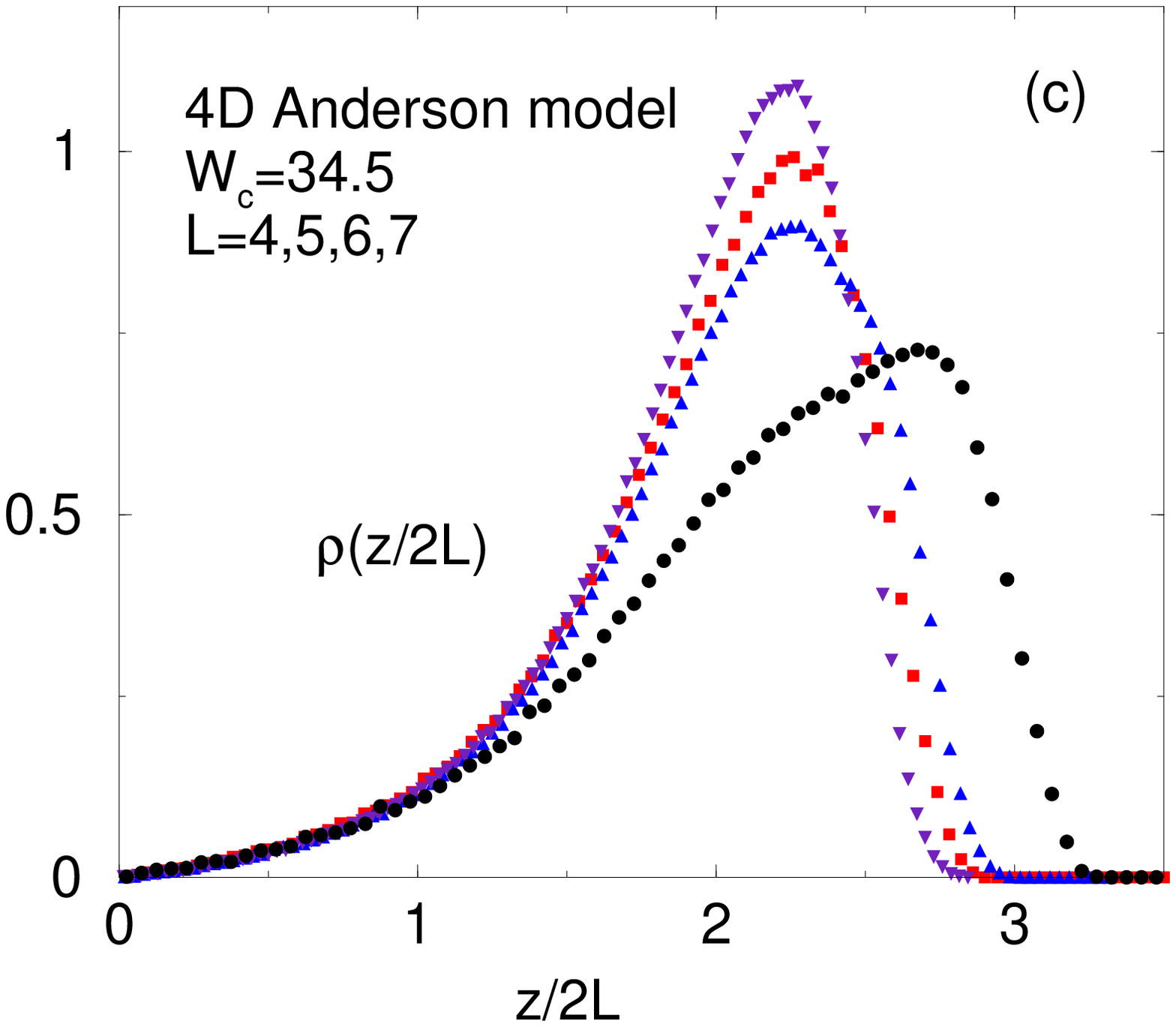}}}
\caption{(a) Critical $\rho(z)$  for different system size. Only small (linear) 
part of $\rho(z)$ is system-size invariant.
(b) normalized  $\rho(z)$ for 3D  and (c) 4D  system.}
 \label{rho_z}
\end{figure}

The critical density  deserves the special Figure. In Fig. 3a
we present $\rho(z)$ for different size of the sample. 
It is evident, that only small part of $\rho(z)$ is system size invariant.
The interval of the size invariance equals  approximately  to that at which
$\rho(z)$ remains linear. When comparing $\rho(z)$ for different
system sizes, $>L_{\rm min}$, then the interval, at which all $\rho$'s coincides, 
is $z\le L_{\rm min}/2$.  This restriction must be taken into account 
in analysis of the scaling properties of higher $z$'s
\cite{condmat}.

In Fig. 3b,c we present the normalized critical density for 
3D and 4D Anderson models. For 4D, $\rho(z)\sim z^2$
in the limit of small $z$. 


The form of $\rho(z)$
remain approximately the same also when $W$ exceeds critical point.
The whole distribution only shifts towards the higher values of $z$.
In the limit of large disorder  we find that
\begin{equation}\label{w_univ}
\rho_{W_1}(z-\langle z_1(W_1)\rangle)=\rho_{W_2}(z-\langle z_1(W_2)\rangle)
\end{equation}
(Fig. 4).
The most remarkable differences could be found only in the tail of the
density for small values of $z$.

\begin{figure}[t]
\centerline{\resizebox{7.1cm}{7.1cm}{\includegraphics{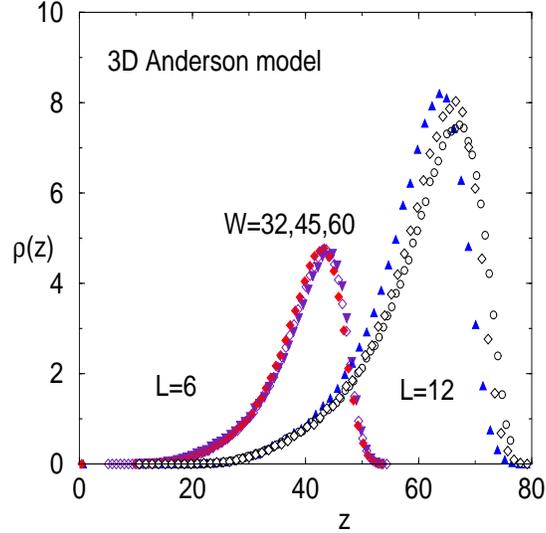}}}
\caption{%
Density  $\rho(z)$ for the 3D system with $W=32$ ($\circ$), 45
($\diamondsuit$)  and 60 (full symbols) for two system size
($L=8$ and 12). We shifted the densities for $W=32$ and $45$ 
in $\langle z_1(W=60)\rangle-\langle  z_1(W)\rangle$ to
show their the universal form
(\ref{w_univ}). 
$\langle z_1\rangle$ for $W=60$ is 20 (3)  for $L=8$ (12), respectively.}
\end{figure}

\smallskip

In conclusion, we present numerical data for Lyapunov
exponents for 3D  and 4D Anderson models in the neighbor of the critical point.
We show how two the most famous characteristics of the statistics of $z$s' change
when the strength of the disorder exceeds the weak disorder limit.
In difference to the level statistics, we found no Semipoisson distribution
of the differences of $z$'s. 
Our data support belief that the statistics of $z$'s  and, consequently,  
of the conductance
has a simple form similar that developed for the metallic regime.

\smallskip

\noindent{Acknowledgment} This work has been supported by Slovak Grant Agency,
Grant n. 2/4109/98. I thank A. von Humboldt Foundation for financial
support.

\end{document}